\documentclass[12pt,preprint,natbib]{aastex}
\shorttitle{Arecibo Observations of 18 Pulsars}
\shortauthors{Lewandowski et al.}
\usepackage{natbib}
\begin{document}

\title{Arecibo Timing and Single Pulse Observations of 18 Pulsars}

\author{Wojciech Lewandowski}
\affil{Toru\'n Centre for Astronomy, Nicolaus Copernicus University, Gagarina 11, 87-100 Toru\'n, Poland}

\author{Alex Wolszczan}
\affil{Department of Astronomy and Astrophysics, The Pennsylvania State University, 525 Davey Laboratory, University Park, PA 16802, USA}
\affil{Toru\'n Centre for Astronomy, Nicolaus Copernicus University, Gagarina 11, 87-100 Toru\'n, Poland}

\author{Gra\. zyna Feiler}
\affil{Toru\'n Centre for Astronomy, Nicolaus Copernicus University, Gagarina 11, 87-100 Toru\'n, Poland}

\author{Maciej Konacki}
\affil{Nicolaus Copernicus Astronomical Center, Polish Academy of Sciences, Rabia\,nska 8, 87-100 Toru\'n, Poland}
\affil{Department of Geological and Planetary Sciences,
       California Institute for Technology, MS 150-21, Pasadena, CA 91125, USA}

\author{Tomasz So\l tysi\' nski}
\affil{Institute of Physics, Szczecin University, Wielkopolska 15, 70-451 Szczecin, Poland}

\begin{abstract}
We present new results of timing and single pulse measurements for 
18 radio pulsars discovered in 1993 - 1997 by
the Penn State/NRL declination-strip survey conducted with the 305-m Arecibo
telescope at 430 MHz. Long-term timing measurements have led
to significant improvements of the rotational
and the astrometric parameters of these sources, including the millisecond pulsar, PSR J1709+2313,
and the pulsar located within the supernova remnant S147, PSR J0538+2817. Single pulse
studies of the brightest objects in the sample have revealed an unusual "bursting" pulsar,
PSR J1752+2359, two new drifting subpulse pulsars, PSR J1649+2533 and PSR J2155+2813, and another
example of a pulsar with profile mode changes, PSR J1746+2540.
PSR J1752+2359 is characterized by bursts of emission, which appear once every 3-5 min. and 
decay exponentially on a $\sim$45 sec timescale. PSR J1649+2533 spends $\sim$30\% of the time
in a null state with no detectable radio emission.
\end{abstract}


\section{Introduction}

Most of the large-scale, undirected pulsar surveys conducted since the time of discovery of
the millisecond pulsars (Backer et al. 1982) have been driven
by an astrophysically well motivated desire to detect more of these rapidly rotating 
neutron stars (e.g. Phinney \& Kulkarni 1994; Lorimer 2001). This is certainly true for
the spectacularly successful southern hemisphere surveys with the Parkes 64-m 
telescope (Manchester et al. 2001; Edwards \& Bailes 2001), 
as well as for a number of less extensive, but more
sensitive Arecibo surveys that have contributed several milestone pulsar discoveries 
(Camilo 1995). In fact, the progress in this field has become particularly
dramatic after the first discoveries of millisecond pulsars away from the
galactic plane (Wolszczan 1991), which indicated an approximately isotropic
distribution of these objects in the solar neighborhood.

Along with the millisecond pulsars, the surveys usually discover the objects of
the so-called slow pulsar variety in much greater numbers. For example, the Arecibo 
drift-scan searches carried out since 1991 have found 9 millisecond pulsars and at 
least 70 slow pulsars (Foster et al. 1995; Camilo et al. 1996; Ray et al. 1996; Xilouris et al. 
2000; McLaughlin, Cordes \& Arzoumanian 2000; Lommen et al. 2000; Lorimer et al. 2002; 
Chandler 2003). Newly discovered slow pulsars
certainly deserve to be subjected to close scrutiny. For example, 
there exists a possibility that some slow pulsars
have black hole or planetary companions (e.g. Thorsett \& Dewey 1993; 
Lipunov et al. 1994; Kalogera et al. 2001). 
Furthermore, timing measurements of pulsars, especially the younger ones, remain
the only source of experimental evidence concerning the neutron star seismology
\citep[see][]{cheng88}. Finally, there is also an open issue of the pulsar emission
mechanism \citep{mel2000}, whose final clarification may require new, 
possibly unanticipated kinds of evidence. 

The Penn State/NRL survey (Foster et al. 1995) and the subsequent 
untargeted observations by the Penn State 
group throughout the Arecibo upgrade period have effectively covered about
850 square degrees of the sky in a manner dictated by the logistics
of the telescope scheduling. They 
have contributed 3 new millisecond pulsars and 16 slow ones to the total number of the drift-scan
survey discoveries quoted above. 
The millisecond pulsars PSR J1640+2224 and PSR J1713+0747 
have been systematically monitored at Arecibo
since the time of their discovery. The timing models for these objects have been published
by Wolszczan et al. (2000) and Camilo, Foster and Wolszczan (1994), respectively. Our
continuing observations discussed in this paper
have established an accurate timing model for the third millisecond
pulsar, PSR J1709+2313, positively verified two of the four unconfirmed slow pulsar candidates
included in Foster et al. (1995), PSR J2151+2315 and PSR 2155+2813, 
and discovered two additional slow pulsars, PSR 1813+1822 and PSR J1908+2351. Two of our
seemingly original discoveries, PSR J1549+21 and PSR J1906+16, 
have proven to be pulsars detected by other searches (Lorimer, private communication; 
Camilo 1995). Finally,
the slow pulsar candidates, PSR B1840+13 and PSR B2208+18, and the millisecond pulsar
candidate, PSR B1735+13, have not been confirmed.

In this paper, we present refined timing models 
for the millisecond pulsar PSR J1709+23, the pulsar PSR J0538+2817 in the supernova remnant S147,
and the sixteen confirmed slow pulsars discovered by the Penn State/NRL surveys since 1991.
The model for PSR J1709+2313 includes a significant proper motion measurement. For all pulsars
listed in Foster et al. (1995), both the spin parameters and the timing positions have been
significantly improved. Timing parameters for the two new slow pulsars are published here
for the first time. We also present the results of single pulse observations of the brightest
objects in the sample. These include the unusual ``bursting'' pulsar, PSR J1752+2359, two new
drifting subpulse pulsars, PSR J1649+2533 and PSR J2155+2813, and a new mode changing pulsar,
PSR J1746+2540.

Our observations and data analysis are described in Section 2. In Section 3, we give
the details of the new timing models for the millisecond pulsar PSR J1709+2313,
PSR J0538+2817, the pulsar located inside the supernova remnant S147,
and for the 16 slow pulsars.
Section 4 is devoted to the single pulse analysis of three pulsars from this sample and
our conclusions are given in Section 5.

\section{Observations and Data Analysis}

The pulse timing and the single pulse observations discussed in this paper have been
made with the 305-m Arecibo radiotelescope using the dual-circular polarization
receiving systems at 430 MHz and 1400 MHz and the Penn State Pulsar Machine (PSPM).
Timing observations made at Arecibo before
1995 with the 40 MHz correlation spectrometer as a pulsar backend are discussed in
Foster et al. (1995) and Cadwell (1997). 

The PSPM pulsar backend is a 2 $\times$ 128 $\times$ 60 kHz,
computer controlled processor designed
to conduct fast sampled pulsar searches and precision timing measurements.
Technical details of the backend
are given in Cadwell (1997). In the timing mode,
the dual-polarization, left- and right-circularly
polarized signals were added together, smoothed with appropriate time
constants ranging from 32 $\mu$s to 700 $\mu$s, 4-bit
quantized, and averaged synchronously with the Doppler-shifted pulsar period.
The average pulse profiles, with the number of phase bins chosen to ensure
an approximately one milliperiod time resolution, were time-tagged using the
observatory's hydrogen-maser clock and stored for further analysis. A typical
integration time for each profile was 180 s. For the purpose of single pulse
analysis, the data were taken in the search mode by continuously sampling the
polarization summed signal at 80 $\mu$s intervals using a 100 $\mu$s post-detection
time constant. In this case, the phasing of the single pulses was accomplished in
the off-line analysis.

The topocentric times-of-arrival (TOA) of pulses were calculated
by cross-correlating the average pulses with high signal-to-noise
template profiles. 
For each pulsar, the standard timing analysis package TEMPO 
(http://www.pulsar.princeton.edu/) was then employed to convert the TOAs to
the solar system barycenter using the JPL DE200 ephemeris (Standish, 1990) 
and to perform least-squares fits of appropriate timing models to the data. 
In the case of slow pulsars,
the models included the pulsar rotation period $P$ and its slowdown rate $\dot P$, 
celestial coordinates $\alpha_{J2000}$ and $\delta_{J2000}$, and
the dispersion measure $DM$. For PSR J1709+2313, the additional
model parameters were the pulsar's proper motion and the five Keplerian elements of its
binary orbit. The proper motion was also included in the timing model for PSR J0538+2817,
in which case the fitting of astrometric parameters was carried out in ecliptic
coordinates, because of the pulsar's proximity to the ecliptic plane (see also
Kramer et al. 2003).

\section{Timing Models and Other Parameters for Individual Sources}

The timing results for two of the 18 pulsars presented in this paper, PSR J1709+2313 and
PSR J0538+2817, are discussed separately in this Section, 
because of their respective memberships
of the millisecond pulsar population and the group of young pulsars associated with
supernova remnants. A joint discussion is presented for
the remaining sources, for which the basic timing model was sufficient
to describe their properties. 

The pulse widths and flux densities
of the 18 pulsars were measured following the procedures described by Camilo and Nice (1995)
and Lorimer, Camilo and Xilouris (2002). The widths listed in Tables 1-3 are given as the 
equivalent width $w_e$ (the width of a top-hat pulse, having the same area and height 
as the profile), $w_{50}$, and $w_{10}$ measured at the 50\%, and 10\% intensity level, 
respectively. Finally, the timing data were used to compute the most important 
derived pulsar parameters (Table 4): 
the distance estimate, $d$, obtained from the Taylor and Cordes (1993) model of the Galactic
electron density distribution; height above the Galactic plane, $z$=$d$ sin $b$; pulsar
luminosity at 430 MHz, $L_{430}$=$S_{430}d^2$; spin-down age, $\tau$=$P/{2\dot P}$;
spin-down energy-loss rate, $\dot E$=$4\pi ^2I\dot P/P^3$ (the moment of inertia
$I$=10$^{45}$ g cm$^2$); and dipole magnetic field, $B$=3.2$\times$10$^{19}(P\dot P)^{1/2}$ G.
Uncertainties of the parameters that involve the pulsar distance depend on
the corresponding errors in distance estimates derived from the Taylor \& Cordes
(1993) model and may be as large as 30\%.

\subsection{The Millisecond Pulsar PSR J1709+2313}

The 4.6-ms pulsar PSR J1709+2313 was discovered in June 1994, during the final months
before the upgrade related shutdown of the Arecibo telescope. The initial timing
model for this faint binary millisecond pulsars has been published by Foster et al. (1995).
Because of the insufficient data, the model could not be significantly improved
until after the reopening of the telescope in late 1997 (Cadwell 1997). 

New observations conducted in 1997-2003 involved systematic
TOA measurements at 430 MHz. Only a few successful 1400 MHz observations were made, due to
the pulsar's faintness at frequencies above 1 GHz. 
The average pulse profiles
of PSR J1709+2313 at the two frequencies (Fig. 1a) are quite similar and exhibit at least three
broad components and an interpulse. 
The two frequency observations
were essential in obtaining an accurate DM measurement of the pulsar. 
They were also used to estimate the pulsar's flux density (Table 1) and the corresponding
spectral index between 430 MHz and 1400 MHz. Because the 1400 MHz flux was measurable
during the infrequent episodes of interstellar
scintillation maxima of the pulsar, 
only an upper limit to the spectral index, $\alpha <<-2$, could
be derived. In fact, its real value is likely to be among the steepest spectral indices measured
for the millisecond pulsars (e.g. Kramer et al. 1998).

The best-fit residuals for
the timing model assuming a 22.7-day, low-eccentricity binary orbit of the pulsar are shown
in Fig. 1b. With the new TOA measurements of PSR J1709+2313, it was possible to obtain
a phase-connected timing solution that spans the entire 1994-2003 observing period and
includes a significant proper motion measurement (Table 1). Because the pulsar is weak
and has a broad pulse profile, the rms residual for the best-fit model is 
relatively large and amounts to 20 $\mu$s for 180-s integrations or 5.5 $\mu$s for daily-averaged
TOAs. 

The very small $\dot P$ of PSR J1709+2313 (Table 1) makes it necessary to correct
its value for kinematic biases related to acceleration due to proper motion
(Shklovskii 1970), and to vertical and differential accelerations in the
Galaxy (Damour \& Taylor 1991). Following these authors, a total kinematic
correction to $\dot P$ takes the form:
\begin{equation}
{\left(\dot P\over P\right)^k} = -{{a_z\sin b}\over c} - {v^2_0\over cR_0}
 \left[\cos l + {\beta\over{{\sin}^2 l + {\beta}^2}}\right] + {\mu}^2{d\over c},
\end{equation}
where $a_z$ is the vertical component of Galactic acceleration, which can be
calculated from a model of the Galactic potential published by Kuijken \& 
Gilmore (1989),
$l$ and $b$ are the pulsar's galactic coordinates, $\mu$ is its proper motion, 
$c$ is the speed of light, $R_0$ and $v_0$ denote the
galactocentric radius and galactic circular velocity of the Sun, respectively, 
and $\beta\equiv d/R_0-\cos l$. For PSR J1709+2313, all the three terms in
equation (1) are significant and yield a total kinematic correction, $\dot P_k$,
listed in Table 1. The error in $\dot P_k$ is dominated by a $\sim$30\% uncertainty
of the pulsar distance estimate. The derived parameters of the pulsar
given in Table 4 have been calculated with the corrected value of $\dot P$.
 
\subsection{PSR J0538+2817 in the Supernova Remnant S147}

PSR J0538+2817, discovered by Anderson et al. (1996), is located within the well-known
supernova remnant S147 (G180.0$-$1.7). Based on the positional coincidence and 
a reasonable similarity of the estimated
distances and ages of the two objects, these authors have hypothesized that the pulsar
could be physically related to the remnant. Initial Arecibo timing observations of PSR J0538+2817 
discussed by Anderson et al. (1996)
were carried out in 1994 and they were continued with the Effelsberg and Jodrell Bank
telescopes by Kramer et al. (2003).

We have resumed timing observations of PSR J0538+2817 with the PSPM after the Arecibo upgrade and
collected TOA measurements at 430 MHz and 1400 MHz over a 2.5-yr period
beginning in late 1998. The average pulse profiles of the pulsar have been published
by Anderson et al. (1996).
Good quality flux density measurements of PSR J0538+2817 at the two frequencies (Table 2) were used
to calculate its spectral index $\alpha=-$1.2$\pm$0.1. This value is the same as the index
derived by Maron et al. (2000) from the flux density measurements at 1.4 GHz and 4.9 GHz and it
implies that the emission of this pulsar over the 0.43 GHz - 4.9 GHz frequency range
is adequately described by a simple power-law spectrum.

A timing model including the pulsar's spin and astrometric
parameters and the dispersion measure was fitted to data in two steps. First,
the two frequency TOA measurements were used to determine DM variations over 
the observing period. To accomplish this, the DM was averaged over five 
partially overlapping, 3-6 month intervals between 1999 and 2002.
As illustrated in Fig. 2a, the DM of PSR J0538+2817 shows an approximately monotonic
increase at a rate of $\sim$0.008 pc cm$^{-3}$/yr. Second, with the DM variations fixed,
the remaining parameters were determined from a least-squares fit of the model to
the TOA measurements. Slow TOA variations caused by the timing noise were fitted out
by including a second time derivative of the pulsar's spin period in the model.
Because of a relatively short, 2.5-yr span of the data used in this analysis, we found
it unnecessary to model the timing noise by means of more precise methods (e.g.
Hobbs, Lyne \& Kramer 2003).
The resulting best-fit TOA residuals are shown in Fig. 2b and the final model parameters
are listed in Table 2. The fit is characterized by an rms residual of 169 $\mu$s
(111 $\mu$s for daily-averaged TOAs). Other pulsar parameters derived from the timing
model are given in Table 4.

\subsection{Results for the 16 Slow Pulsars}

Approximate timing models for 12 out of the 16 ``ordinary'', slow pulsars discussed
here have been published by Foster et al. (1995) along with the list of unconfirmed
pulsar candidates. Our timing observations initiated after the Arecibo upgrade have
resulted in significant improvements of these models and in establishing the models
for two more confirmed pulsars from the original list, PSR J2151+2315 and PSR J2155+2813.
In addition, sufficient data have been gathered for two more recently discovered
pulsars, PSR J1813+1822 and PSR J1908+2351, to determine their timing behavior for
the first time.

All the post-upgrade timing observations of the 16 pulsars were conducted at 430 MHz
between late 1998 and mid-2001. For pulsars observed before 1995, the new TOAs were
phased with the old ones across the 3-yr Arecibo upgrade gap in the model fitting
process. For the four newer objects mentioned above, the TOA modeling was based on
the data collected over a 2.5-yr post-upgrade period. The least-squares fits of
the timing models included the standard spin and astrometric parameters of the pulsars.
Because no second frequency data were available for these objects, their
dispersion measures were determined by splitting the 8 MHz bandpass of the PSPM into
two bands, 4 MHz apart, calculating the TOAs for the two center frequencies and
fitting for DM using the two-frequency TOA sets.
The best-fit timing and derived parameters for all the 16 pulsars 
are listed in Tables 3 and 4, respectively. Typically,
the models fitted have rms residuals in the 0.5 - 1.5 ms range for a 180 s integration time
per TOA measurement, depending on the pulse strength and width. 

The average pulse profiles of 12 pulsars discovered before 1995 have been published
by Foster et al. (1995). The profiles of the four newer pulsars discussed above are
shown in Fig. 3. Also shown are the two profiles of PSR J1746+2540 which was found
to undergo clearly distinguishable profile mode changes.
In addition, we display in Fig. 3 an improved profile for PSR J1938+0650, whose shape
originally published by Foster et al. (1995) was artificially broadened,
because of the incorrect timing model used in the pulse averaging process.

\section{Single Pulse Behavior of Three Pulsars}

The purpose of the single pulse analysis was to search the data for possible presence
of phenomena such as subpulse drifting, nulling or pulse mode
changes. To accomplish this, blocks of 1024 to 4096
consecutive single pulses were phased up
using the ephemerides derived from the up-to-date pulse timing models
and displayed as pulse phase vs. pulse number diagrams for visual inspection
and further analysis.

\subsection{Drifting Subpulses in PSR J1649+2533 and PSR J2155+2813}

Drifting subpulses represent an organized behavior of the emission peaks within
the average pulse envelope which manifests itself as a correlated progression
of the subpulse phase over consecutive pulse periods (e.g. Backer 1973).
Another phenomenon, more common in older pulsars, is the pulse nulling - a sudden
cessation of the pulsed emission followed by an equally sudden emission onset several
periods later (e.g. Ritchings 1976, Biggs 1992).

We have identified the drifting subpulses in PSR J1649+2533 and PSR J2155+2813.
Because the drift rate is high in both pulsars and cannot be easily detected in
the raw single pulse data, the corresponding drift patterns are presented in Fig. 4
in the form of two-dimensional autocorrelation functions of the pulse intensity
distribution in the pulse number/pulse phase plane. These plots clearly reveal
the presence of periodic patterns of a rapidly drifting emission in both pulsars.

The subpulse drift patterns are usually parametrized in terms of the subpulse separation, $P_2$,
and the spacing between the adjacent subpulse bands, $P_3$ (e.g. Backer 1973).
The measured values of these parameters for PSR J1649+2533 and PSR J2155+2813 are
19.8 ms and 15.7 ms for $P_2$, and 2.2$P$ and 2.5$P$ for $P_3$, respectively. In addition
to subpulse drifting, PSR J1649+2533 exhibits the pulse nulling, which occurs approximately
30\% of the time (Fig. 5a).

\subsection{Burst-like Emission of PSR J1752+2359}

This pulsar has been selected for additional observations
because of its unusually long nulling periods. After the
initial, post-discovery study by Cadwell (1997), we have observed it
on several occasions between 2000 and 2002. As shown in Fig. 5b,
the pulsar
spends 70-80\% of the time in a ``quasi-null'' state. The ``on-states''
occur once every 400-600 periods and last, on average, for $\sim$100 periods.
A more detailed inspection of the on-states (Fig. 5c) reveals that
these burst-like emissions
are quite similar in shape and duration
and they decay into a null state in a manner that is quite
reasonably described by a $te^{-t/\tau}$ function with $\tau\sim$45 seconds.

A comparison of the single pulse characteristics of PSR J1752+2359 with a more typical
nulling behavior of PSR J1649+2533 (Fig. 5a) clearly shows that the two phenomena exhibit
distinctly different morphologies, because
PSR J1752+2359 switches off gradually, rather than suddenly, as observed
in the nulling pulsars (Ritchings 1976, Deich et al. 1986, Biggs 1992). 
Moreover, the astonishing morphological similarity of the individual
on-states and their very similar timescales and repetition rates are unique
among the nulling pulsars.
To our knowledge, these emission characteristics of PSR J1752+2359 make
it an exceptional member of the pulsar population. 

\section{Discussion}

The pulsars discussed in this paper fall into three evolutionary categories as indicated
in the $P$-$\dot P$ diagram shown in Fig. 6. When interpreted in terms of the rotating
magnetic dipole model (Pacini 1968; Gold 1968), the location of a pulsar in this diagram 
depends on the strength of its surface magnetic field ($\sim (P\dot P)^{1/2}$) and on 
its characteristic age ($\tau=P/2\dot P$). Within this framework, 
the relatively young pulsar PSR J0538+2817, possibly
still associated with the supernova remnant S147, appears close to the cluster of
similarly young objects, whereas the 16 ``normal'' pulsars fall in the area that is
occupied by the largest, intermediate age category. The two possible exceptions in this
group are PSR J1813+1822 and PSR J1908+2351 (Table 4), which have characteristic ages
well above 100 Myr, magnetic fields below 10$^{11}$ Gauss, and may more properly be 
categorized as mildly ``recycled''
pulsars which originated from disrupted binaries (e.g. Bailes 1989). 
Finally, as expected, the millisecond
pulsar PSR J1709+2313 is found among other members of the old, recycled pulsar
population (Phinney \& Kulkarni 1994).

\subsection{PSR J1709+2313}

The timing model of PSR J1709+2313 presented here replaces a provisional set of
parameters derived from initial observations of the pulsar conducted just before
the Arecibo upgrade (Foster et al. 1995).
Timing parameters of the pulsar and the resulting old characteristic age,
low companion mass, and weak surface magnetic field (Table 1), place
this object in the class of low-mass binary millisecond pulsars (LMBPs), presumably
orbited by He white dwarf companions (Phinney \& Kulkarni 1994). 
The pulsar's $\sim$90 km s$^{-1}$ transverse
velocity (Table 1) is quite typical for the LMBP population 
(see Cordes \& Chernoff, 1997). 

A common explanation for the unrealistically old spindown age of the pulsar
is that it must have been born
with a spin period close to its observed value 
(e.g. Camilo, Thorsett \& Kulkarni 1994; Nice \& Taylor 1995).
Assuming that the observed, kinematically corrected 
spindown, $\dot P_i$, is entirely driven by a magnetic dipole
radiation, and requiring that the pulsar cannot be older than the estimated
age of the Galaxy ($\sim 10\times 10^9$ yr; Winget et al. 1987),
one can estimate the initial spin period, $P_0$, from the equation:
\begin{equation}
\tau = {P\over 2\dot P_i}{\left[1 - \left(P_0\over P\right)^2\right]},
\end{equation}
which reduces to its simpler, more frequently used form, when $P_0\ll P$.
In the case of PSR J1709+2313, equation (2) yields $P_0\geq$4.1 ms indicating very
little period evolution for this pulsar.

Another interesting feature of PSR J1709+2313 is
its 22.7-day orbital period, which falls in the ``period gap'' invoked
in the context of modeling of the paths of binary evolution that may lead up to formation
of the LMBPs (Camilo 1995). Another such case, PSR J1618-39 with a 22.8-day binary period,
has been detected by Edwards and Bailes (2001). As remarked by these authors,
the existence of the two objects does weaken the period gap case, but it still does not
remove a possibility of underpopulation of the orbital period range between 12 days and
56 days.

\subsection{PSR J0538+2817}

PSR J0538+2817 in the supernova remnant S147 has been recently studied by Kramer et al. (2003). 
These authors have used a proper
motion measurement of the pulsar, based on timing observations with the Jodrell Bank
and the Effelsberg telescopes, to conclude that the object moves away from the center of
the remnant at $\sim$385 km s$^{-1}$. This result implies that the pulsar's characteristic
age of $\sim$0.6 Myr dramatically overestimates its true age, which now appears to
be close to 30 kyr, and that the pulsar had to be born with the initial spin period not much shorter
than the observed one. 

The Arecibo timing parameters of PSR J0538+2813 (Table 2) agree
with those obtained by Kramer et al. (2003) within the measurement errors. Although
the significance of our proper motion estimate for the pulsar is low, because of the short, 2.5-yr
span of observations, both the magnitudes and signs of this effect in the ecliptic coordinates
are in a reasonable agreement with the values published by Kramer et al. Therefore, we
conclude that, within the precision limits imposed by the extent of the Arecibo timing data,
they are consistent with the results presented by these authors in favor of
the actual association of PSR J0538+2813 with S147.

\subsection{PSR J1649+2533, PSR J1752+2359, and PSR J2155+2813}

The slow pulsars, PSR J1649+2533 and PSR J2155+2813 belong to the slowly growing
group of relatively rare objects characterized by drifting subpulses, nulling and,
in some cases, a combination of both (see Ord, Edwards \& Bailes 2001, Lorimer, Camilo \&
Xilouris 2002, Chandler 2003, and Edwards \& Stappers 2003, 
for the description of other recent detections).
Studies of these phenomena have been regarded as very important for our understanding of
the detailed physics of pulsar emission (e.g. Ruderman \& Sutherland 1975, Biggs 1992,
Deshpande \& Rankin 1999). This appears to be particularly true for analyses of
the interaction between subpulse drifting and nulling (Lyne \& Ashworth 1983,
Deich et al. 1986, Vivekanand \& Joshi 1997, van Leeuwen et al. 2003). As no unified
picture of this interaction has emerged so far, possibly because of the insufficient
statistics, new pulsars like PSR J1649+2533, displaying both drifting and nulling,
are obvious candidates for further, more detailed observations.

The spin characteristics of the bursting pulsar, PSR J1752+2359 make it
a fairly typical member of the slow pulsar population (Table 3, Fig. 6).
In addition, the timing and the
pulse profile morphology data do not indicate any pulse arrival time and/or
pulse shape variability that might be due to orbital motion or precessional
beam wobble and could be used to interpret the observed
exponentially decaying bursts of emission from the pulsar (Fig. 5b,c).
As the pulsar's unusual properties may provide new, useful constraints on the emission
process, it is important to observe it at multiple frequencies, including
polarization measurements, to achieve a possibly complete phenomenological
description of its radio signal before
attempting any generalizing conclusions.
It may also be interesting to explore an intriguing
possibility that PSR J1752+2359 may be a member of a sizeable group of pulsars
of the same type that have remained undiscovered because of the very
low duty cycle, burst-like behavior of such objects.
\acknowledgments

WL, MK, GF and AW were partially  supported from the KBN grant 2.P03D.006.16. AW and WL
acknowledge partial support from the NSF grant AST-9988217.
The authors thank B. Cadwell and R. Foster for their contributions to the early
phases of this research.
The Arecibo observatory is part of the National Astronomy and Ionosphere center, which
is operated by Cornell University under a cooperative agreement with the National 
Science Foundation.

\clearpage

\begin{figure}
\plotone{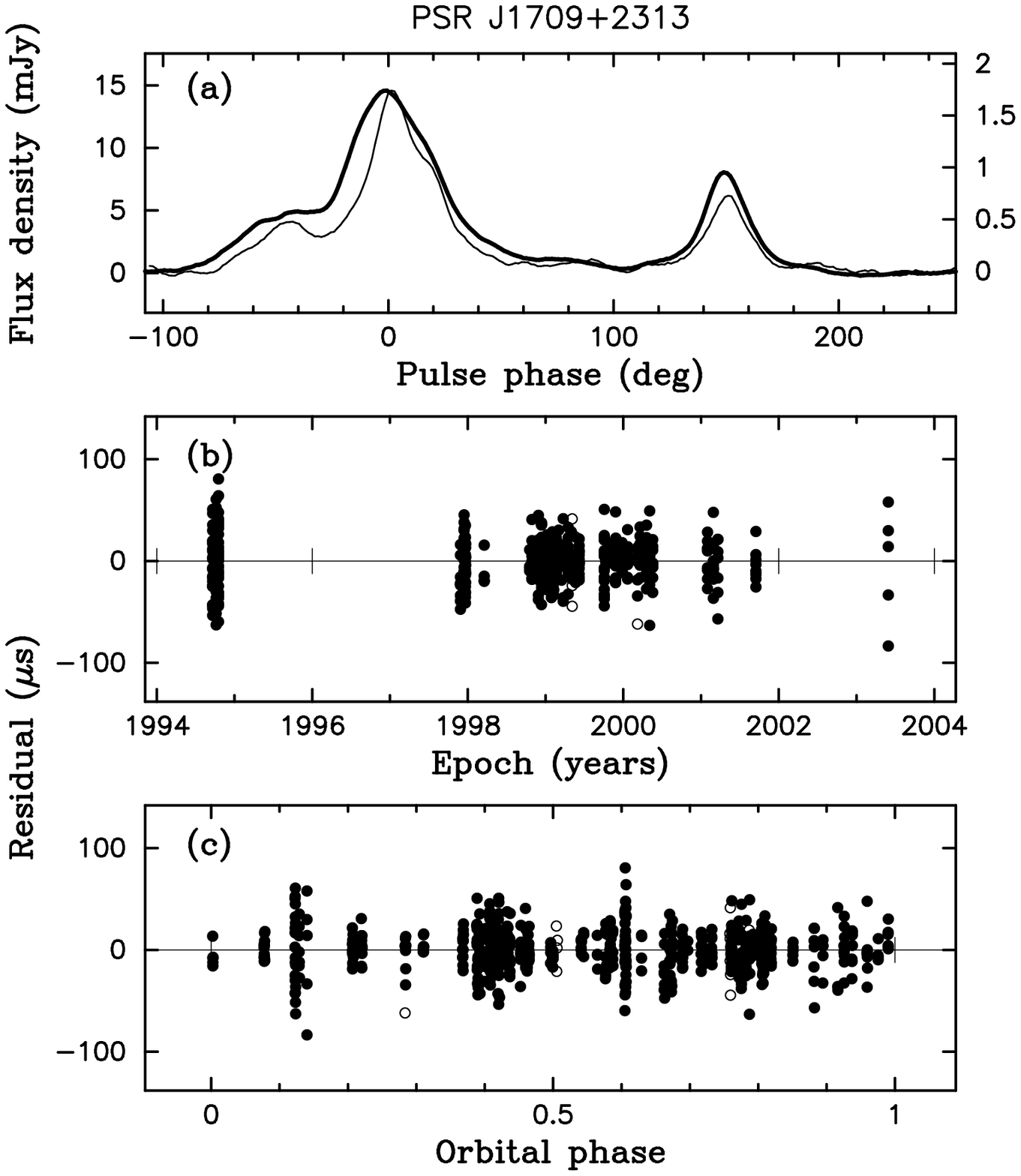}
\caption{Observations of the 4.6-ms pulsar PSR J1709+2313 with the Arecibo telescope.
(a) Average pulse profiles at 430 MHz (heavy solid line) and 1400 MHz (solid line). (b) The best-fit
timing residuals as a function of time at 430 MHz (filled circles) and 1400 MHz (open
circles). (c) Residuals as in (b) plotted against the orbital phase. 
\label{fig1}}
\end{figure}

\clearpage 

\begin{figure}
\plotone{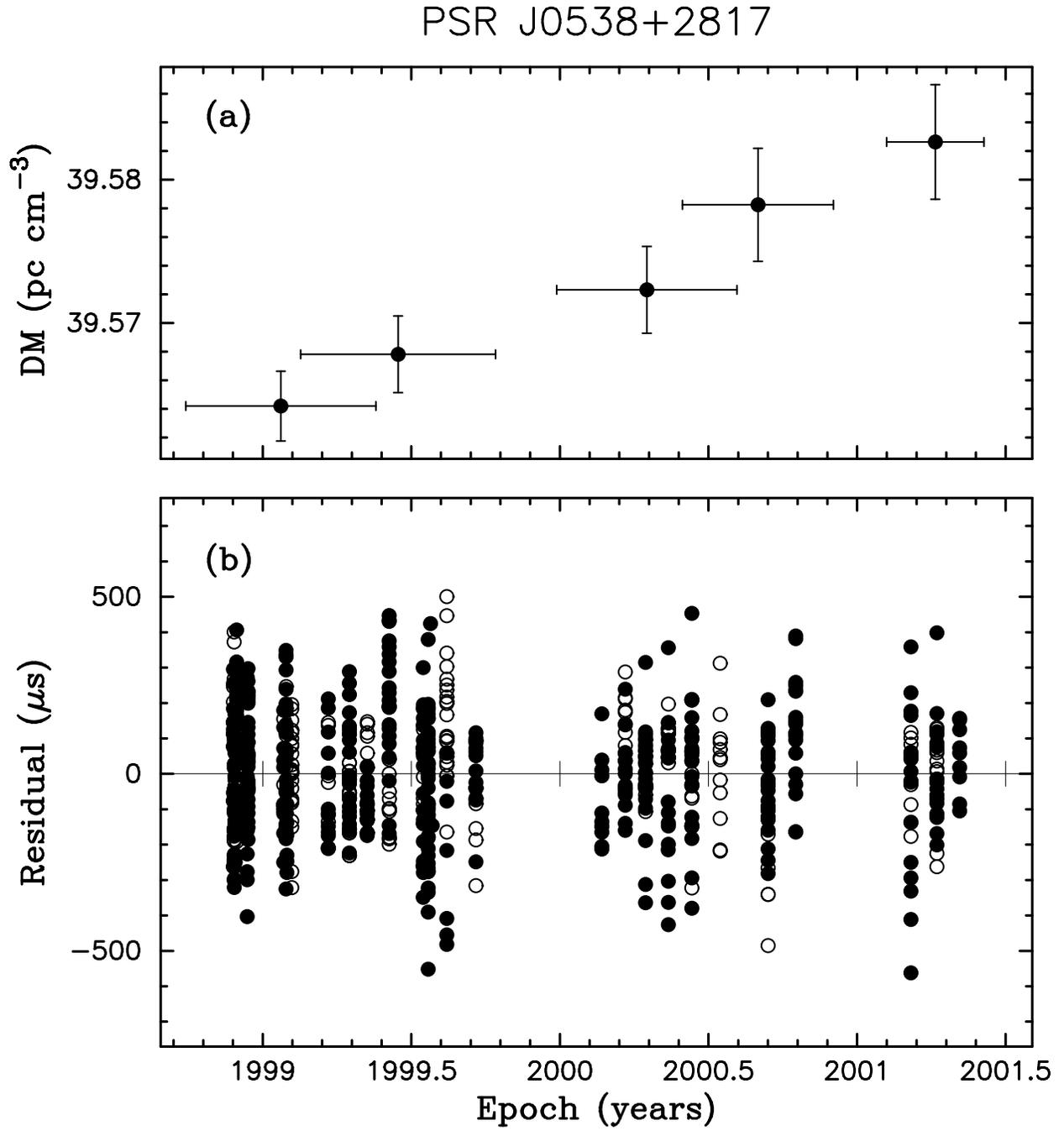}
\caption{Arecibo observations of PSR J0538+2817 in the supernova remnant S147. (a) Dispersion
measure variations over the observing period (see text for details). (b) The best-fit timing
residuals at 430 MHz (filled circles) and 1400 MHz (open circles).
\label{fig2}}

\end{figure}

\clearpage

\begin{figure}
\plotone{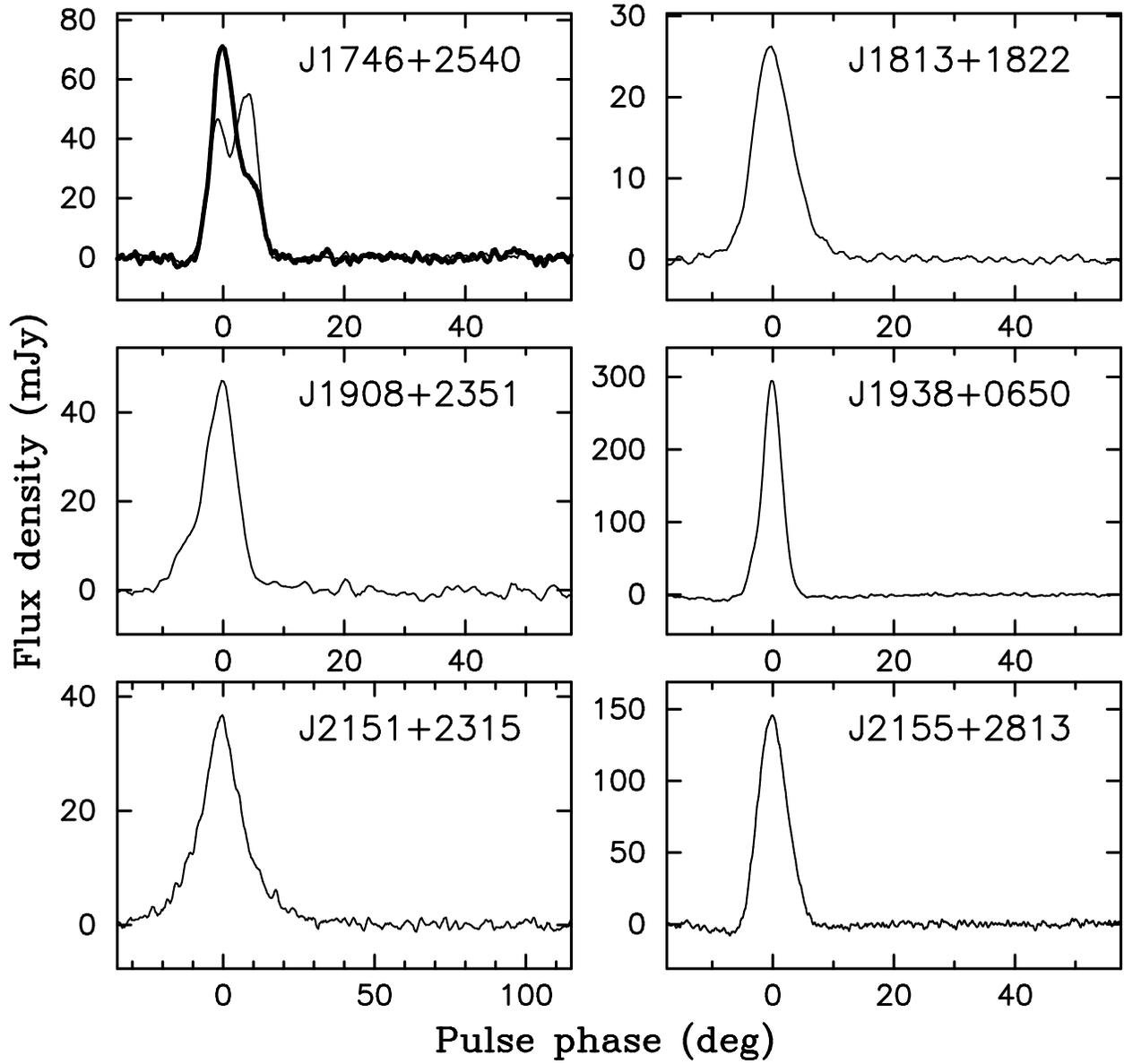}
\caption{Previously unpublished average pulse profiles at 430 MHz. For the mode-changing 
pulsar PSR J1746+2540, two stable profile modes are displayed.
\label{fig3}}
\end{figure}

\clearpage 

\begin{figure}
\plotone{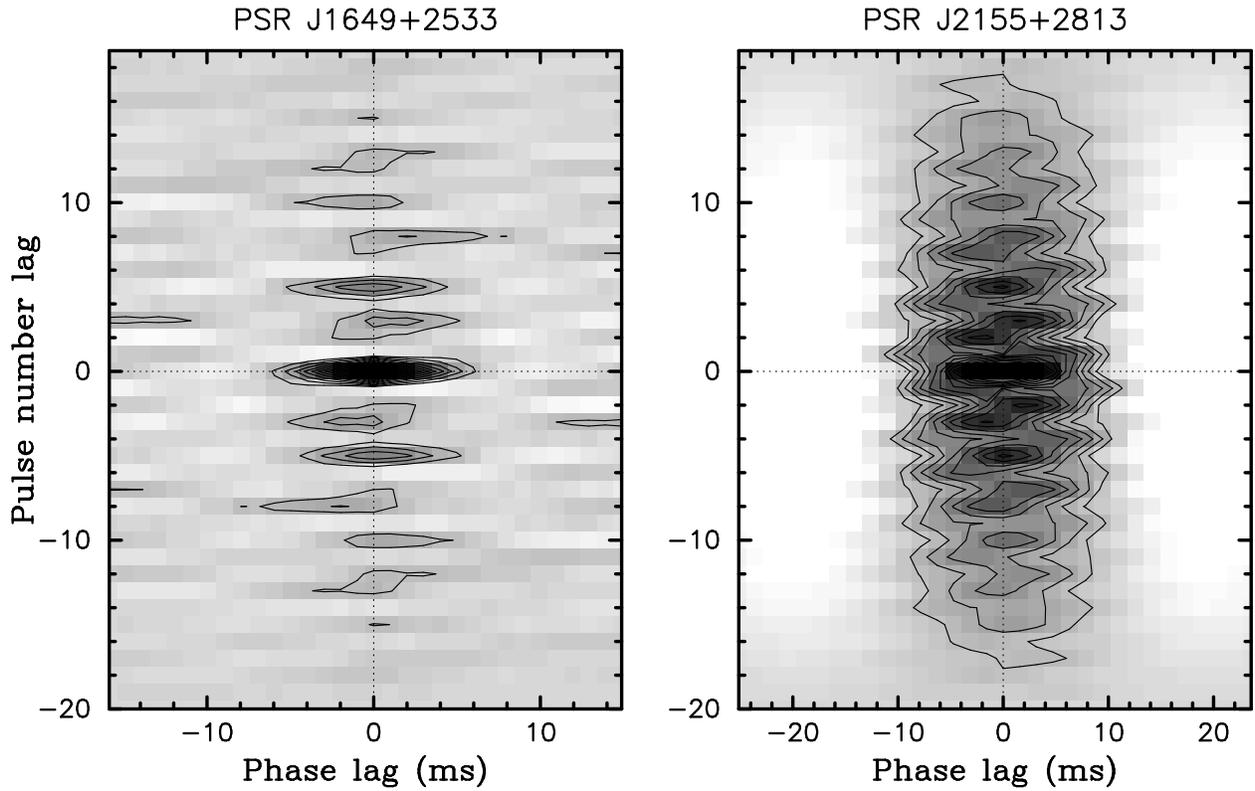}
\caption{Normalized, two-dimensional autocorrelation functions of single pulse sequences from
PSR J1649+2533 and PSR J2155+2813 at 430 MHz. The correlation
coefficient is mapped in the range of 0.05 to 0.95 at the contour interval of 0.05.
\label{fig4}}
\end{figure}

\clearpage 

\begin{figure}
\epsscale{0.8}
\plotone{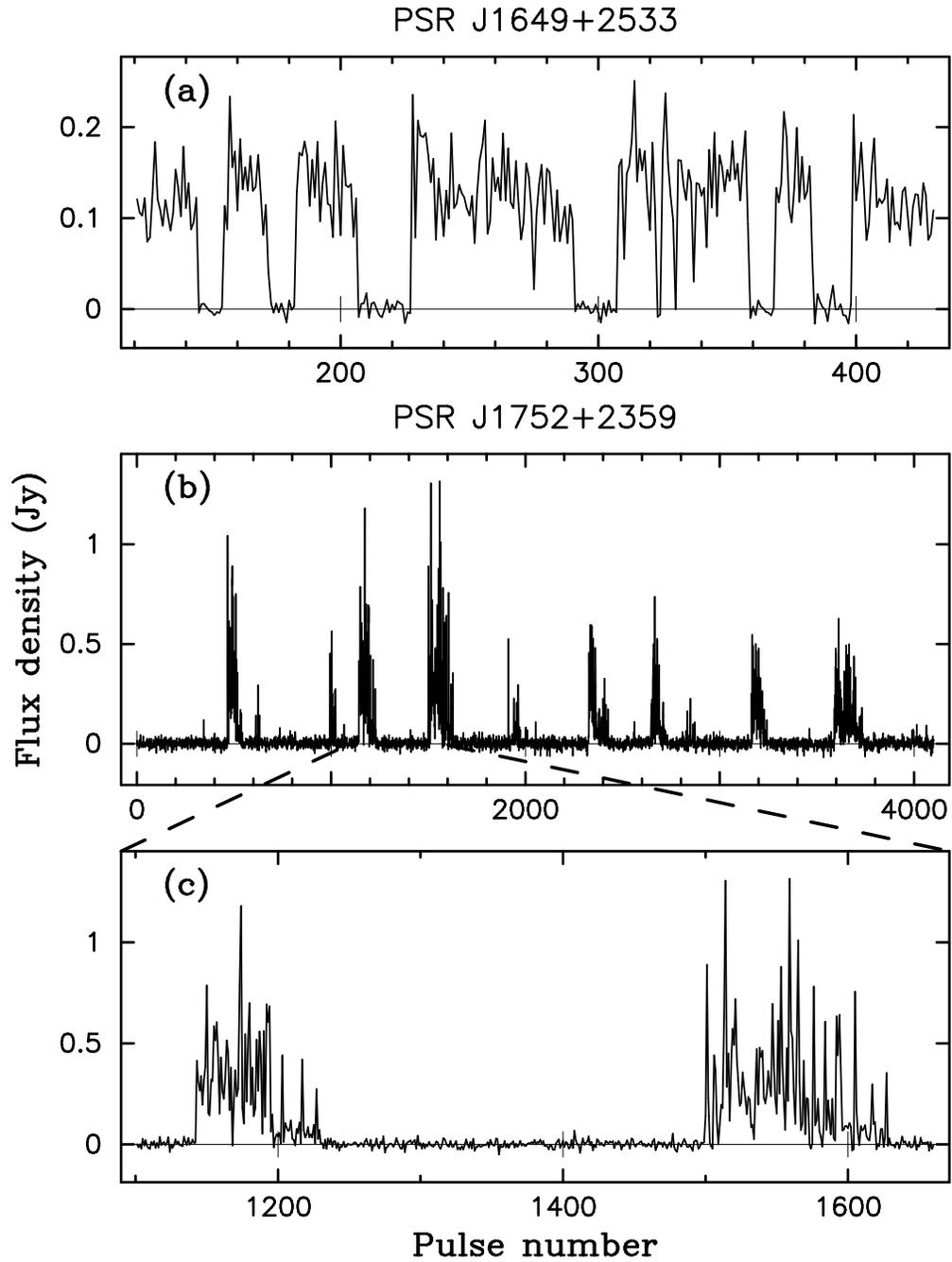}
\caption{Pulse-to-pulse intensity variations in PSR J1649+2533 and PSR J1752+2359 at 430 MHz.
(a) Pulse nulling in PSR J1649+2533. (b) An example of the bursting behavior of
PSR J1752+2359. (c) Details of the intensity variations in two consecutive bursts.
\label{fig5}}
\end{figure}

\begin{figure}
\plotone{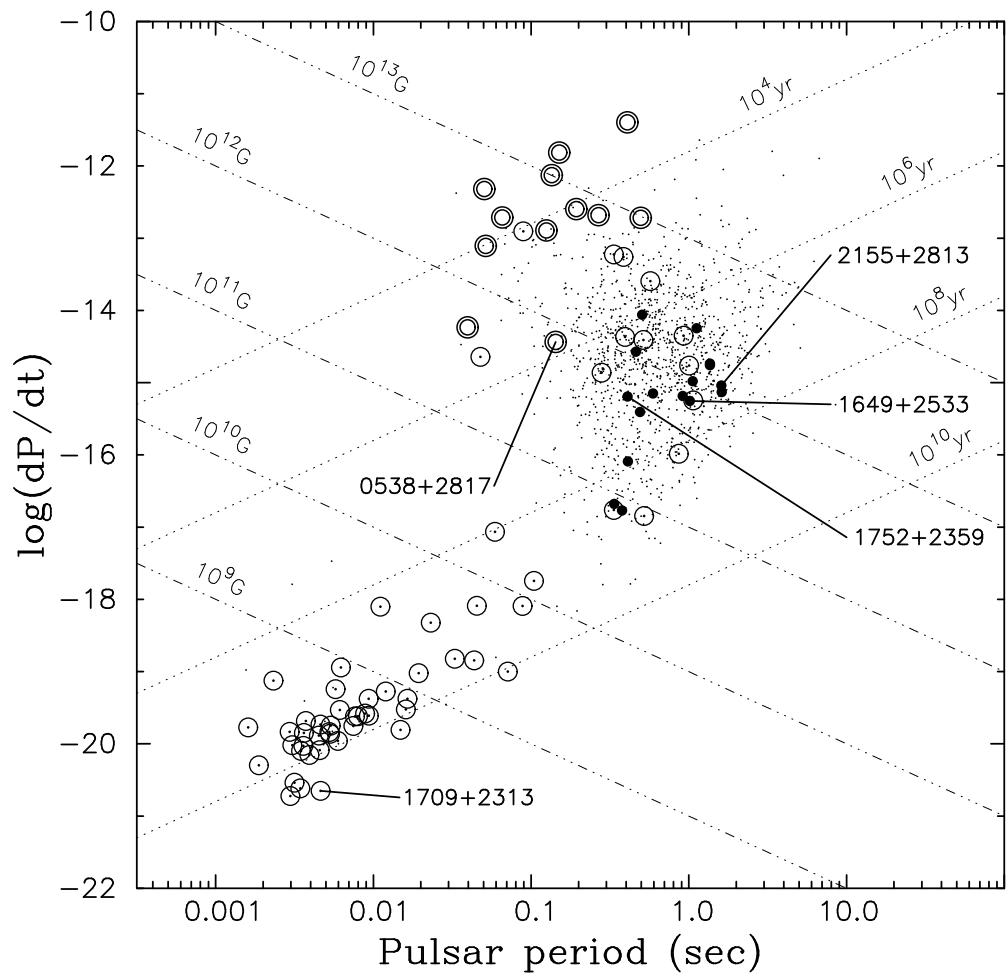}
\caption{Distribution of the 18 pulsars discussed in the text 
in the $P$-$\dot P$ diagram based on the ATNF pulsar catalog 
({\tt http://www.atnf.csiro.au/research/pulsar/catalogue}).
Individually discussed objects are indicated by names. The other
new pulsars are denoted by filled circles. 
The catalog data used
include isolated pulsars (dots), binary pulsars (encircled dots) and pulsars associated
with supernova remnants (double circles).
Lines of constant magnetic field and age are
drawn with the dash-dotted and dotted lines, respectively. 
\label{fig6}}
\end{figure}

\clearpage

\begin{deluxetable}{ll}
\tablecaption{Parameters for PSR~J1709+2313. \label{tab1}}
\tabletypesize{\footnotesize}
\tablewidth{0pt}
\tablehead{
\colhead{Parameter} & \colhead{Value}
}
\startdata
Right ascension $\alpha$ (J2000) \dotfill &     $17^h09^m05\fs792(6)$ \\
Declination $\delta$ (J2000)   \dotfill &     $23^o13'27\farcs85(1)$ \\
Proper motion $\mu_{\alpha}$ (mas yr$^{-1}$) \dotfill &  $-$3.2(7) \\
Proper motion $\mu_{\delta}$ (mas yr$^{-1}$) \dotfill &  $-$9.7(9) \\
Spin frequency $\nu$ (s$^{-1}$) \dotfill &  215.926931187247(7)  \\
First derivative $\dot{\nu}$ ($\times 10^{-15}$ s$^{-2}$)  \dotfill &   $-$0.1692(2) \\
Spin period $P$ (s) \dotfill & 0.0046311962778409(2) \\
Period derivative $\dot{P}$ ($\times 10^{-15}$) \dotfill &  0.00000363(4) \\ 
Kinematic correction $\dot{P_k}$ ($\times 10^{-15}$) \dotfill &  0.0000014(4) \\ 
Position and frequency epoch (MJD)  \dotfill &         51145.0 \\
Dispersion Measure DM (pc cm$^{-3}$) \dotfill & 25.3474(2) \\ 
Orbital period $P_b$ (days) \dotfill & 22.71189238(2) \\ 
Projected semi-major axis $a \sin i$ (lt-s) \dotfill & 15.288544(2) \\
Eccentricity $e$  \dotfill& 0.0000187(2) \\
Time of periastron passage $T_p$ (MJD)\tablenotemark{a} \dotfill & $51196.24079\pm0.036$ \\
Longitude of periastron  $\omega$ (deg)\tablenotemark{a}\dotfill & $24\fdg30300\pm0.58$ \\
Minimum companion mass $m_c (M_{\sun})$ \dotfill & 0.27 \\
Tranverse velocity $v_t$ (km/s) \dotfill & 89(30)\tablenotemark{b} \\
Pulse width at FWHM, $w_{50}$ (ms)\tablenotemark{c}  \dotfill &  0.52 / 0.34 \\
Pulse width at 10\% peak, $w_{10}$ (ms)\tablenotemark{c}  \dotfill& 1.31  / 1.14\\
Equivalent width, $w_e$ (ms)\tablenotemark{c} \dotfill & 0.58 / 0.50 \\
Average flux (mJy)\tablenotemark{c}  \dotfill & 2.52(7) / 0.2(1) \\
\enddata
\tablenotetext{a}{ as the precission of $T_p$ and $\omega$ is too low for 
observational purposes, we quote extra digits for the benefit of future observers}
\tablenotetext{b}{ error includes contribution from uncertainty in distance estimate}
\tablenotetext{c}{ values at 430/1400~MHz}
\end{deluxetable}

\clearpage

\begin{deluxetable}{ll}
\tablecaption{Astrometric, spin and derived parameters for PSR~J0538+2817. \label{tab2}}
\tabletypesize{\footnotesize}
\tablewidth{0pt}
\tablehead{
\colhead{Parameter} & \colhead{Value}
}
\startdata
Ecliptic longlitude $\lambda$ \dotfill & 85\fdg232546(1) \\
Ecliptic latitude $\beta$ \dotfill &  4\fdg93582(3) \\
Proper motion $\mu_{\lambda}$ (mas yr$^{-1}$) \dotfill &  $-$27(6) \\
Proper motion $\mu_{\beta}$ (mas yr$^{-1}$) \dotfill & 129(118)   \\
Right ascension $\alpha$ (J2000)\tablenotemark{a} \dotfill & $5^h38^m25\fs06$ \\
Declination $\delta$ (J2000)\tablenotemark{a}  \dotfill & $28^o17'09\farcs10$ \\
Spin frequency $\nu$ (s$^{-1}$) \dotfill &  6.985269943305(6)  \\
First derivative $\dot{\nu}$ ($\times 10^{-15}$ s$^{-2}$)  \dotfill &  $-$179.0720(6) \\
Spin period $P$ (s) \dotfill & 0.143158390172(1) \\
Period derivative $\dot{P}$ ($\times 10^{-15}$) \dotfill &  3.6699(1) \\ 
Position and frequency epoch (MJD)  \dotfill & 51500.0 \\
Dispersion Measure DM (pc cm$^{-3}$)\tablenotemark{b}  \dotfill & 39.569 \\
Pulse width at FWHM, $w_{50}$ (ms)\tablenotemark{c}  \dotfill & 4.1 / 2.7  \\
Pulse width at 10\% peak, $w_{10}$ (ms)\tablenotemark{c} \dotfill &  17.8 / 13.9 \\
Equivalent width, $w_e$ (ms)\tablenotemark{c} \dotfill & 7.6 / 5.9 \\
Average flux (mJy)\tablenotemark{c}  \dotfill & 8.2(2) / 1.9(1) \\
\enddata
\tablenotetext{a}{ calculated from the best-fit ecliptic coordinates}
\tablenotetext{b}{ see text and Fig.~\ref{fig2} for data on DM variability}
\tablenotetext{c}{ values at 430/1400~MHz}
\end{deluxetable}

\clearpage

\begin{table}
\caption{Parameters for 16 slow pulsars. \label{tab3}}
\vskip5mm
\hspace*{-10mm}{\scriptsize
\begin{tabular}{lllllcccccc}
\tableline
  & \phn\phn\phn\phn R. A. & \phn\phn\phn\phn Dec & \phn\phn\phn\phn $P$ & \phn\phn\phn $\dot{P}$ &Epoch & DM & $w_e$& $w_{50 }$ & $w_{10}$ & $S_{430}$ \\
 \phn\phn\phn PSR & \phn\phn\phn (J2000) & \phn\phn\phn (J2000) & \phn\phn\phn\phn(s) & $(10^{-15})$ & (MJD) & (pc/cm$^3$) & (ms) & (ms) & (ms) & (mJy) \\
\tableline \tableline
J1627+1419 & 16:27:18.77(1) &  14:19:20.5(3) &  0.4908568275(6) &   0.3931(1) & 51825.0 & 33.8(6) & 28 & 30 & 41 & 6.1(3) \\
J1645+1012 & 16:45:34.4(2)   &  10:12:16.0(4) &  0.410860739492(2) &   0.0812(5)& 51825.0 & 36.3(6) & 11 & 11 & 19 & 2.3(1) \\
J1649+2533 & 16:49:44.23(5) &  25:33:07.0(2) & 1.0152573918(5) &   0.5594(2)& 51825.0 & 35.5(9) & 24 & 25. & 38 & 7.4(4) \\
J1652+2651\tablenotemark{a} & 16:52:03.07(3) &  26:51:40.4(4) & 0.91580349720(5) &   0.6537(1) &  51739.0 & 41.2(3) & 24 & 12,14 & 43 & 11.3(1) \\
J1720+2150\tablenotemark{a} & 17:20:01.30 (2) &  21:50:12.8(4) &  1.61566378034(7) &   0.740(2) & 51796.0& 41.1(4) & 25 & 8,7 & 70 & 2.4(1) \\
J1741+2758\tablenotemark{b} & 17:41:53.51(2) &  27:58:09.0(9) & 1.3607376877(4) &   1.841(1) & 51797.0 &29.3(6)& 18 & 7,4 & 40 & 3.0(1) \\
J1746+2540\tablenotemark{A} & 17:46:06.87(6) &  25:40:37.5(1) & 1.0581481703(1) &   1.047(2) & 51796.0 &51.5(2) & 21 & 24 & 33 & 1.2(1) \\
J1746+2540\tablenotemark{B} &                &                &                 &            &         &         & 17 & 13 & 31 & \\
J1752+2359 & 17:52:35.42(2) &  23:59:48.2(2) &  0.409050865044(9) & 0.6427(9) & 51952.0 & 36.0(9)& 5  & 4  & 11 & 3.5(3) \\
J1811+0702\tablenotemark{c} & 18:11:20.42(7) &  07:02:29.7(2) & 0.46171267661(3) &   2.692(1) & 51885.0 & 57.8(1) &17 &  $-$,13 & 15,21 & 2.2(1) \\ 
J1813+1822 & 18:13:38.76(5) &  18:22:15.0(9) &  0.3364246764(9) &   0.021(8) & 51886.0 & 60.8(5)& 8 & 6 & 13 & 0.6(1) \\
J1822+0705 \tablenotemark{d}& 18:22:18.44(2) &  07:05:20.1(1) &  1.3628174508(2) &   1.749(5) &51730.0 & 61.2(5)&21 & 9,10 & 40,19 & 3.8(1)  \\
J1848+0647 & 18:48:56.01(3) &   06:47:31.7(4) & 0.5059567391(2) &   8.7516(1) & 51824.0 & 27.9(2)& 14 & 15 & 27 & 2.3(1) \\
J1908+2351 & 19:08:31.94(7) &  23:51:41.9(8) & 0.377578026(1) &   0.017(1) &51561.0 & 101.5(7)& 7 & 6 & 14 & 0.9(2) \\
J1938+0650 & 19:37:53.46(4) &  06:50:06.0(2) &  1.121561892(5)&   5.68(1) & 51824.0&70.8(2)& 12 & 10 & 22 & 3.2(1) \\
J2151+2315 & 21:51:28.9(1) &  23:15:12.8(5) &  0.593533613(3) &   0.708(3)& 51594.0& 23.6(2)& 30 & 23 & 68 & 1.6(1) \\
J2155+2813 & 21:55:15.82(1) &  28:13:12.1(2) & 1.6090199964(6) &   0.9164(9) & 51594.0& 77.4(2)&25 & 24 & 42 & 2.1(1)\\ \tableline
\end{tabular}}
\tablenotetext{A,B}{ modes of PSR~J1746+2540}
\tablenotetext{a}{ double profile with ``bridge''}
\tablenotetext{b}{ blended multiple component profile, two components rise above 50\% of maximum height}
\tablenotetext{c}{ two separate components, left one does not reach 50\% of maximum height}
\tablenotetext{d}{ two separate components}
\end{table}

\clearpage

\begin{table}
\caption{Derived parameters for 18 pulsars. \label{tab4}}
\vskip5mm
{\small
\begin{tabular}{crccccc}
\tableline
 &     &  & log [$L_{430}$ &  & log$[\dot{E}$ & log \\
Pulsar & $d$(kpc) & $z$(kpc) & (\mbox{mJy kpc}$^2)]$ & $\tau$(Myr) &  (ergs s$^{-1}$)] & $[B(G)]$ \\ \tableline \tableline
J0538+2817..... & 1.78    & 0.052   & 3.23 & 0.62 & 34.7 & 11.9 \\
J1627+1419..... & $>$2.84 & $>$1.76 & 1.7 & 19.7  & 32.1 & 11.6\\
J1645+1012..... & $>$3.27 & $>$1.75 & 1.3 & 80.1  & 31.6 & 11.2\\
J1649+2533..... & $>$2.91 & $>$1.75 & 1.8 & 28.7  & 31.3 & 11.9\\
J1652+2651..... & $>$2.93 & $>$1.76 & 2.0 & 22.2  & 31.5 & 11.9\\
J1709+2313..... & 1.83    & 0.97    & 2.13 & 33300 & 32.4 & 8.01 \\
J1720+2150..... & $>$3.59 & $>$1.76 & 1.5 & 34.6  & 30.8 & 12.0 \\
J1741+2758..... & 2.07  & 0.93 & 0.6 & 11.7  & 31.4 & 12.2\\
J1746+2540..... & $>$4.16 & $>$1.76 & 1.3 & 16.0  & 31.5 & 12.0\\
J1752+2359..... & 2.70  & 1.05 & 1.4 & 10.1  & 32.6 & 11.7\\
J1811+0702..... & 3.13  & 0.65 & 1.3 & 2.7   & 33.0 & 12.0\\
J1813+1822..... & $>$6.24 & $>$1.76 & 1.4 & 253.8 & 31.3 & 10.9\\
J1822+0705..... & 2.98  & 0.50 & 1.5 & 12.3  & 31.4 & 12.2 \\
J1848+0647..... & 1.44  & 0.09 & 0.7 & 0.9   & 31.4 & 12.3 \\
J1908+2351..... & 6.62  & 0.81 & 1.6 & 358.8 & 31.1 & 10.9 \\ 
J1938+0650..... & 3.64  & $-$0.45 & 1.6 & 3.1  & 32.2 & 12.4 \\
J2151+2315..... & 1.42  & $-$0.57 & 0.5 & 13.2 & 32.1 & 11.8 \\
J2155+2813..... & $>$5.06 & $<-$1.76 & 1.73 & 27.8 & 30.9 & 12.1\\ \tableline
\end{tabular}
}
\end{table}
 

\begin{thebibliography}{}
\bibitem[Anderson et al. (1996)]{and96}Anderson, S. B., Cadwell, B. J., Jacoby, B. A., Wolszczan, A., Foster, R. S. \& Kramer, M. 1996, \apj, 468, 55
\bibitem[Backer(1973)]{backer73} Backer, D.~C 1973, \apj, 182, 245
\bibitem[Backer  et al. (1982)]{backer82} Backer, D.~C., Kulkarni, S.~R., Heiles, C., Davis, M.~M. \& Goss, W.~M.  1982, \nat, 300, 615
\bibitem[Bailes(1989)]{bailes1989} Bailes, M. 1989, \apj, 342, 917
\bibitem[Biggs(1992)]{biggs92} Biggs, J.~D. 1992, \apj, 394, 574
\bibitem[Cadwell(1997)]{cadwell97} Cadwell, B. 1997, PhD Thesis, Pennsylvania State University
\bibitem[Camilo, Foster \& Wolszczan(1994)]{camilo94} Camilo, F., Foster, R. S. \& Wolszczan, A. 1994, \apjl, 437, L39
\bibitem[Camilo(1995)]{camilo95} Camilo, F. 1995 , in {The Lives of the Neutron Stars,  Alpar, M. A., Kiziloglu, U., van Paradijs, J. (eds) (NATO ASI Series C, 515; Dordrecht: Kluwer)}, 243
\bibitem[Camilo \& Nice(1995)]{cam_nice95} Camilo, F., Nice, D. J. 1995, \apjl, 445, 756
\bibitem[Camilo et al.(1996)]{camilo96} Camilo, F., Nice, D. J., Shrauner, J. A., Taylor \& J. H. 1996, \apj, 469, 819  
\bibitem[Chandler(2003)]{chandler2003} Chandler, A. 2003, PhD Thesis, California Institute of Technology
\bibitem[Cheng et al.(1988)]{cheng88} Cheng, K. S., Pines, D., Alpar, M. A. \& Shaham, J. 1988, \apj, 330, 835
\bibitem[Cordes \& Chernoff(1997)]{cord97} Cordes, J.~M. \& Chernoff, D.F. 1997, \apj, 482, 971
\bibitem[Damour & Taylor(1991)]{damour1991} Damour, T. \& Taylor, J. H. 1991, \apj, 366, 501
\bibitem[Deich et al.(1986)]{deich86} Deich, W. T. S., Cordes, J. M., Hankins, T. H. \&  Rankin, J. M. 1986, \apj, 300, 540
\bibitem[Deshpande \& Rankin(1999)]{desh1999} Deshpande, A. A. \& Rankin, J. M. 1999, \apj, 524, 1008
\bibitem[Edwards \& Bailes(2001)]{edwards2001} Edwards, R. T., Bailes, M. 2001, \apj, 553, 801
\bibitem[Edwards \& Stappers(2003)]{edwards2003} Edwards, R. T. \& Stappers, B. W. 2003,  \aap, 407, 273 
\bibitem[Foster et al.(1995)]{fos95} Foster, R. S., Cadwell, B. J., Wolszczan, A. \& Anderson, S. B. 1995, \apj, 454, 826
\bibitem[Gold(1968)]{gold1968} Gold, T. 1968, \nat, 218, 731
\bibitem[Hobbs, Lyne \& Kramer(2003)]{hobbs2003} Hobbs, G., Lyne, A.~G. \& Kramer, M. 2003, to appear in proceedings of the conference "Radio Pulsars" held in Chania, Crete, Sept 2002
\bibitem[Kalogera at al.(2001)]{kalogera2001} Kalogera, V., Narayan, R., Spergel, D. N. \& Taylor, J. H. 2001, \apj, 556, 340 
\bibitem[Kramer at al.(1998)]{kramer1998} Kramer, M., Xilouris, K. M., Lorimer, D. R., Doroshenko, O., Jessner, A., Wielebinski, R.,  Wolszczan, A. \& Camilo, F. 1998, \apj, 501, 270
\bibitem[Kramer et al.(2003)]{kramer2003} Kramer, M., Lyne, A.~G., Hobbs, G., L\"{o}hmer, O., Jordan, C. \& Wolszczan, A. 2003, submitted to ApJL 
\bibitem[Kuijken \& Gilmore(1989)]{kuijken1989} kuijken, K. \& Gilmore, G. 1989, \mnras, 239, 605
\bibitem[van Leeuwen et al.(2003)]{vanleeuwen2003} van Leeuwen, A. G. J., Stappers, B. W., Ramachandran, R. \& Rankin, J. M. 2003, \aap, 399, 223
\bibitem[Lipunov et al.(1994)]{lipunov94} Lipunov, V. M., Postnov, K. A., Prokhorov, M. E., Osminkin, E. Yu. 1994, \apjl, 423, L121
\bibitem[Lommen et.al(2000)]{lommen2000} Lommen, A. N., Zepka, A., Backer, D.~C., McLaughlin, M., Cordes, J.~M.,  Arzoumanian, Z. \& Xilouris, K. 2000, \apj, 545, 1007
\bibitem[Lorimer(2001)]{lorimer2001} Lorimer, D.~R. 2001, Living Reviews in Relativity, vol. 4, art. 5 
\bibitem[Lorimer, Camilo \& Xilouris(2002)]{lorimer2002} Lorimer, D.~R., Camilo, F. \& Xilouris, K.~M. 2002, \aj, 123, 1750
\bibitem[Lyne \& Ashworth(1983)]{lyne83} Lyne, A. G. \& Ashworth, M. 1983, \mnras, 204, 519
\bibitem[Manchester et al.(2001)]{manchester2001} Manchester, R. N., Lyne, A. G., Camilo, F., Bell, J. F., Kaspi, V. M., D'Amico, N., McKay, N. P. F., Crawford, F., Stairs, I. H., Possenti, A., Kramer, M. \& Sheppard, D. C. 2001, \mnras, 328, 17
\bibitem[Maron et al.(2000)]{maron2000} Maron, O., Kijak, J., Kramer, M. \& Wielebinski, R. 2000, \aap Sup., 147, 195
\bibitem[McLaughlin, Cordes \& Arzoumanian(2000)]{mclaugh2000} McLaughlin, M. A., Cordes, J. M., Arzoumanian, Z. 2000, in {\sl IAU Coll. 177, Pulsar Astronomy - 2000 and Beyond, Kramer, M.,  Wex, N. \&  Wielebinski, R. (eds) (San Fransisco: ASP)}, 41 
\bibitem[Melrose(2000)]{mel2000} Melrose, D. B. 2000, in {\sl IAU Coll. 177, Pulsar Astronomy - 2000 and Beyond, Kramer, M.,  Wex, N. \&  Wielebinski, R. (eds) (San Fransisco: ASP)}, 721
\bibitem[Ord, Edwards \& Bailes(2001)]{ord2001} Ord, S. M., Edwards, R. \&  Bailes, M. 2001, \mnras, 328, 911
\bibitem[Pacini(1968)]{pacini68} Pacini, F. 1967, \nat, 219, 145
\bibitem[Phinney \& Kulkarni(1994)]{phinney94} Phinney, E.~S. \&  Kulkarni, S.~R. 1994, \araa, 32, 591
\bibitem[Ray et~al.(1996)]{ray96} Ray, P. S., Thorsett, S. E., Jenet, F. A., van Kerkwijk, M. H., Kulkarni, S. R., Prince, T. A., Sandhu, J. S. \& Nice, D. J., \apj, 470, 1103
\bibitem[Ritchings(1976)]{ritchings76} Ritchings, R. T. 1976, \mnras, 176, 249
\bibitem[Ruderman \&  Sutherland(1975)]{ruder75} Ruderman, M. A. \& Sutherland, P. G. \apj, 196, 51
\bibitem[Shklovskii(1970)]{shklovski1970} Shklovskii, I.~S. 1970, Soviet Astron., 13, 562
\bibitem[Standish(1990)]{standish1990} Standish, E.~M.~Jr. 1990, \aap, 233, 252
\bibitem[Taylor \& Cordes(1993)]{taylor93} Taylor, J. H. \& Cordes, J. M. 1993, \apj, 411, 674
\bibitem[Thorsett \& Dewey(1993)]{thorsett93} Thorsett, S. E., Dewey, R. J. 1993, \apjl, 419, 65
\bibitem[Vivekanand \& Joshi(1997)]{vive1997} Vivekanand, M. \& Joshi, B. C. 1997, \apj, 477, 431
\bibitem[Winget et al.(1987)]{winget1987} Winget, D. E., Hansen, C. J., Liebert, J., van Horn, H. M., Fontaine, G., Nather, R. E., Kepler, S. O. \& Lamb, D. Q. 1987, \apj, 315, L77
\bibitem[Wolszczan(1991)]{wol1991} Wolszczan, A. 1991, \baas, 23, 834
\bibitem[Wolszczan et al.(2000)]{wol2000} Wolszczan, A., Doroshenko, O., Konacki, M., Kramer, M., Jessner, A., Wielebinski, R., Camilo, F., Nice, D. J., \& Taylor, J. H. 2000, \apj, 528, 907
\bibitem[Xilouris (2000)]{xilouris2000} Xilouris, K. M., Fruchter, A., Lorimer, D. R., Eder, J. \& Vazquez, A. 2000, in {\sl IAU Coll. 177, Pulsar Astronomy - 2000 and Beyond, Kramer, M.,  Wex, N. \&  Wielebinski, R. (eds) (San Fransisco: ASP)}, 21
\end{thebibliography}
\end{document}